\documentclass[pdflatex,sn-mathphys]{sn-jnl}

\usepackage[font=footnotesize,labelfont=bf,justification=justified,format=plain]{caption}
\usepackage[T1]{fontenc}

\theoremstyle{thmstyleone}%
\newtheorem{theorem}{Theorem}
\newtheorem{proposition}[theorem]{Proposition}%

\theoremstyle{thmstyletwo}%

\theoremstyle{thmstylethree}%

\usepackage{etoolbox}   
\makeatletter
\patchcmd{\@maketitle}{\artauthors}{\centering{\artauthors}}{}{}
\makeatother

\begin{document}

\begin{figure}[t]
\includegraphics[width=0.3\textwidth]{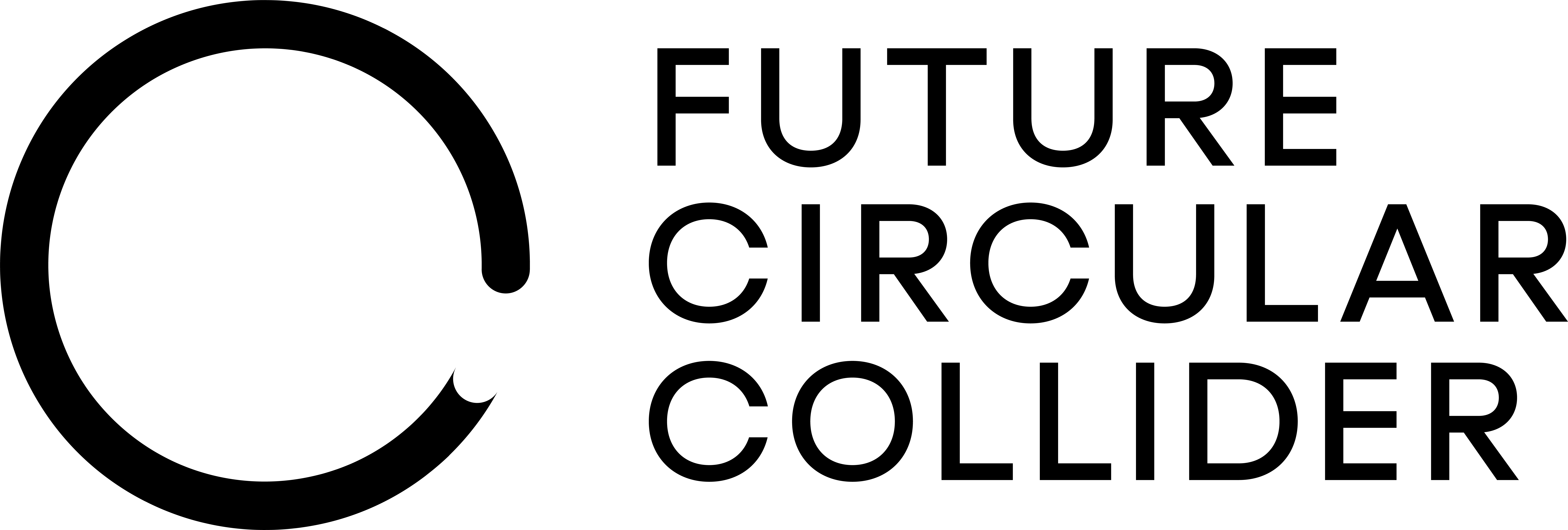}
\end{figure}


\title{The carbon footprint of proposed $\rm e^+e^-$ Higgs factories}


\author*[1]{\fnm{Patrick} \sur{JANOT}}
\author[2,3]{\fnm{Alain} \sur{BLONDEL}}



\affil[1]{\small \orgname{CERN}, \orgaddress{\street{1 Esplanade des Particules}, \postcode{1217} \city{Meyrin}, \country{Switzerland}}}
\affil[2]{\small \orgname{LPNHE, IN2P3-CNRS}, \orgaddress{\street{4 place Jussieu}, \postcode{75005} \city{Paris}, \country{France}}}
\affil[3]{\small \orgname{University of Geneva}, \orgaddress{\street{24 rue du G\'en\'eral-Dufour}, \postcode{1211} \city{Geneva}, \country{Switzerland}}}
\affil[*]{\small e-mail: \href{mailto:patrick.janot@cern.ch}{patrick.janot@cern.ch}}





\abstract{The energy consumption of any of the $\rm e^+e^-$ Higgs factory projects that can credibly operate immediately after the end of LHC, namely three linear colliders (CLIC, operating at $\sqrt{s}\,=\,380$\,GeV; and ILC and $\rm C^3$, operating at $\sqrt{s}\,=\,250$\,GeV) and two circular colliders (CEPC and FCC-ee, operating at $\sqrt{s}\,=\,240$\,GeV), will be everything but negligible. Future Higgs boson studies may therefore have a significant environmental impact. This note proposes to include the carbon footprint for a given physics performance as a top-level gauge for the design optimization and, eventually, the choice of the future facility. The projected footprints per Higgs boson produced, evaluated using the 2021 carbon emission of available electricity, are found to vary by a factor 100 depending on the considered Higgs factory project.}




\maketitle

\newpage
\section{Introduction}\label{section:introduction}

In January 2020, the European Strategy Group concluded that an electron-positron Higgs factory is the highest-priority next collider, thus giving to CERN the mandate to prepare for a Higgs factory, followed by a future hadron collider with sensitivity to energy scales an order of magnitude higher than those of the LHC~\cite{CERN-ESU-015}. The CERN Council endorsed this vision in June 2020. In its June 2021 session, the CERN Council approved~\cite{FCC-FS-plans:202106,FCC-FS-org:202106} and funded~\cite{CERNmtp} the FCC technical and financial feasibility study, with focus on the first step, i.e., the tunnel and the first stage machine FCC-ee. This additional step somehow clarified the path for Europe towards the intermediate future.

In July 2022, the Snowmass'21 effort held its last public meeting in Seattle~\cite{Snowmass21}, and the Energy Frontier working group came to similar general conclusions~\cite{Reina}, pending the P5 recommendations and the subsequent DoE decision: after the immediate future (HL-LHC), the intermediate future is an $\rm e^+e^-$ Higgs factory, based on either a linear (ILC in Japan~\cite{ILCInternationalDevelopmentTeam:2022izu}, $\rm C^3$ in the US~\cite{Dasu:2022nux}, CLIC at CERN~\cite{Brunner:2022usy}) or a circular (FCC-ee at CERN~\cite{Bernardi:2022hny}, CEPC in China~\cite{Gao:2022lew}) collider, and the long-term future is a multi-TeV ($\mu\mu$ and/or pp) collider. 

The choice between three linear colliders and two circular colliders may appear, if only scientific arguments are considered, somewhat subjective. Indeed, the extensive discussions of the scientific pros and cons throughout the European Strategy and Snomwass studies led to the conclusion that a substantial part of the Higgs physics programmes is common to linear and circular $\rm e^+e^-$ colliders. In addition, significant domains of physics exist, in which each of these two types of machines is unique and irreplaceable~\cite{Blondel:2019ykp}: 
\begin{itemize}
    \item Linear colliders have the ability to explore lepton collisions above 400\,GeV. This is particularly interesting for the searches for new particles in the gaps possibly left by hadron colliders, and may prove essential should a new particle be discovered at the LHC in the suitable energy regime. In addition, centre-of-mass energies above 1\,Tev also give the opportunity of a first (often called ``direct'') measurement of the Higgs self-coupling via double Higgs production, with an expected precision of about 10\%. Above about 2\,TeV, muon colliders are generally considered more effective. 
    \vskip .15cm
    \item For circular colliders, the high luminosity and the multiple detectors, the exquisite energy calibration at the Z, WW, ZH and $\rm t\bar t$ energies, and the possibility of monochromatisation at $\sqrt{s} = m_{\rm H}$, are building blocks of a unique program. The Z factory with several trillions of Z produced, offers a multitude of electroweak and QCD measurements, flavour (b, c, $\tau$) physics, as well as direct searches for SM symmetry violations and feebly coupled particles. The possibility to observe the $s$-channel Higgs production leads to a unique chance of measuring the Yukawa coupling of the electron. A circular collider also remains the most pragmatic, safest, and most effective way towards a 100\,TeV pp exploration machine.
\end{itemize}
Surely, the members of the linear (circular) Higgs factory efforts are keen on putting forward that, should a circular (linear) factory go ahead without a complementary linear (circular) collider, considerable physics opportunities might be lost. 

In any case, the acceptance of the next facility by the governments and the public will be highly facilitated if it can be shown that environmental concerns have taken an important place in the choice. In our times of measurable climate change due to a global warming of unprecedented rapidity, resulting from the emission of greenhouse gases, a most important criterion is that the scientific outcome for a given carbon footprint be maximised.

\section{Carbon footprint per physics outcome} \label{section:tie-breaker}

The Higgs factory actors have made no mistake about it: they regularly communicate about their potential energy savings while in operation with better technology choices and sustained technology R\&D, as well as about the recycling and recovery of as much as possible of the energy that cannot be saved in the first place (e.g., under the form of domestic heating, etc.). Several statements can be found in the literature on this important topic, which do not immediately help towards a resolution of the dispute. Below are three statements that lead to completely different propositions: 
\vspace{-0.5cm}
\begin{proposition}
The estimated electrical power for the linear Higgs factories is in general lower than circular machines~\cite{forum}.
\end{proposition}
\vspace{-1.2cm}
\begin{proposition}
The ILC is certainly the less “energy hungry” of all future projects thanks, in particular, to its superconducting radiofrequency (RF) technology~\cite{ILCnewsline}.
\end{proposition}
\vspace{-1.2cm}
\begin{proposition}
The expected AC site power consumption for Higgs production, calibrated to performance, is probably the lowest for the FCC-ee, compared with other proposed future projects~\cite{Benedikt:2020ejr}. 
\end{proposition}
\vspace{-0.7cm}
To reconcile the three propositions and draw conclusions as to the environmental footprint of each collider, it is necessary to go through the actual numbers. The third row of Table~\ref{tab:power}, which displays the instantaneous power consumption estimates ($P$) of the five Higgs factory projects (CLIC at 380\,GeV, ILC/$\rm C^3$ at 250\,GeV, and CEPC/FCC-ee at 240\,GeV) as reported in Ref.~\cite{Roser:2022sht},  validates the first proposition. The annual energy consumption in operation, $E$, is calculated by multiplying the instantaneous power by the time spent in collisions $T$, to which is added for completeness the product of the time spent with the collider powered but not in collision mode ($=T/\epsilon-T$, where $\epsilon$ is the operational efficiency) by the instantaneous power in that mode, assumed to be half of the full power. This seemed to validate the second proposition, at least until the recent CLIC performance update~\cite{Brunner:2022usy} was made public. 

\begin{table}[htbp]
\caption{For each of the Higgs factory projects (1st row): Centre-of-mass energy (2nd row); Instantaneous wall-plug power~\cite{forum} (3rd row); Assumed annual operational time and operational efficiency~\cite{Bordry:2018gri,Bambade:2019fyw,
Blondel:2019yqr,CEPCStudyGroup:2018rmc,CEPCPhysicsStudyGroup:2022uwl} (4th and 5th rows); Inferred annual energy consumption in operation (last row).}
\label{tab:power}
\begin{center}
\begin{tabular}{@{}lccccc@{}}
\toprule
Higgs factory & CLIC & ILC & $\rm C^3$ & CEPC & FCC-ee\\
$\sqrt{s}$ (GeV) & 380 & 250 & 250 & 240 & 240 \\
\midrule
Instantaneous power $P$ (MW) & 110 & 140 & 150 & 340 & 290 \\
Annual collision time $T$ ($10^7$ s) & 1.20 & 1.60 & 1.60 & 1.30 & 1.08 \\
Operational efficiency $\epsilon$ (\%) & 75 & 75 & 75 & 60 & 75 \\
Annual energy consumption $E$ (TWh) & 0.4 & 0.7 & 0.8 & 1.6 & 1.0 \\
\botrule
\end{tabular}
\end{center}
\end{table}

As this point, it is important to ask whether or not the instantaneous power and the annual energy consumption are relevant measures for the environmental impact of a given factory. Maybe not! Take for example a car factory that produces 1000 (electric) cars per year, consuming energy $E_1$, and a second factory producing 10,000 such cars every year with twice the total energy $E_2 (=2E_1)$. If an environment-aware firm wants to buy 5,000 cars, they will naturally choose the second factory, which can produce these cars with five times less energy, not to  mention the additional bonus that the totality of the cars will be delivered ten times quicker. Therefore, one concludes that to meet the client's demands of a given number of cars, the latter factory is five times more environmental-friendly than the former, in spite of its twice larger annual energy consumption. The same is true for a Higgs factory: given that the physics outcome depends on the total number of Higgs bosons produced, a Higgs factory must be judged by its ability to produce Higgs bosons, and in particular by the energy needed to produce each Higgs boson, if the environmental impact is of any concern.  

The number of Higgs bosons and the time needed to produce them, as expected in the current operation models of the five Higgs factories, are shown in Table~\ref{tab:Higgs}. The last row displays the energy consumption per Higgs boson as obtained by combining these figures with the last row of Table~\ref{tab:power}. The FCC-ee is 
the most energy-efficient Higgs factory, with 3\,MWh/Higgs: this observation validates the third proposition above.  
\begin{table}[htbp]
\caption{For each of the Higgs factory projects (1st row): Running time in the current operation model (2nd row); Total number of Higgs boson produced with the baseline values for the centre-of-mass energy, beam longitudinal polarisation and integrated luminosity (3rd row); Energy consumption per Higgs boson (4th row). }
\label{tab:Higgs}
\begin{center}
\begin{tabular}{@{}lccccc@{}}
\toprule
Higgs factory & CLIC & ILC & $\rm C^3$ & CEPC & FCC-ee\\
\midrule
Running time as a Higgs factory (year) & 8 & 11.5 & 11.5 & 10 & 3 \\
Total number of Higgs bosons produced ($10^6$) & 0.25 & 0.5 & 0.5 & 4 & 1 \\
\midrule
Energy consumption per Higgs boson (MWh) & 14 & 17 & 18 & 4.1 & {\bf 3.0} \\
\botrule
\end{tabular}
\end{center}
\end{table}
The baseline FCC-ee operation model has so far included only two interaction points (IP), but the collider layout was recently modified to be compatible with the operation of up to four detectors~\cite{Bernardi:2022hny}. The total number of Higgs bosons expected to be produced at these four IPs is expected to be up to 1.7 times larger than that produced at two IPs, thus reducing the energy consumption per Higgs boson to less than 1.8\,MWh/Higgs. As a Higgs factory, {\bf FCC-ee with four IPs would therefore be one order-of-magnitude more energy-efficient than ILC or $\bf C^3$}, for example. 

While the energy efficiency expressed in MWh/Higgs is a closer evaluation of the environmental footprint of each collider than the instantaneous power or the annual energy consumption, it still does not give the full and accurate 
picture. The environmental impact is dominated by the contribution to global warming resulting from  greenhouse gas emission, which itself strongly depends on where and when the energy is produced. Today, more than 90\% of the electricity used to operate the CERN accelerators originates from carbon-free sources (nuclear, hydro-electric, solar, and wind) in France. As a consequence, a MegaWatt-hour used at CERN to power a collider has a much smaller carbon footprint than a MegaWatt-hour used in the US, in Japan, or in China.

\begin{figure}[htbp]%
\centering
\begin{sideways}
\begin{minipage}{19.5cm}
\caption{Carbon intensity of electricity production around the world in 2021, in $\rm kgCO_2$ equivalent / MWh, obtained from Ref.~\cite{CO2}. The carbon footprints of the energy produced in Illinois, Japan, and France, and used respectively at FNAL, KEK, and CERN, are highlighted.}\label{fig:CO2}
\vspace{0.2cm}
\includegraphics[width=1.0\textwidth]{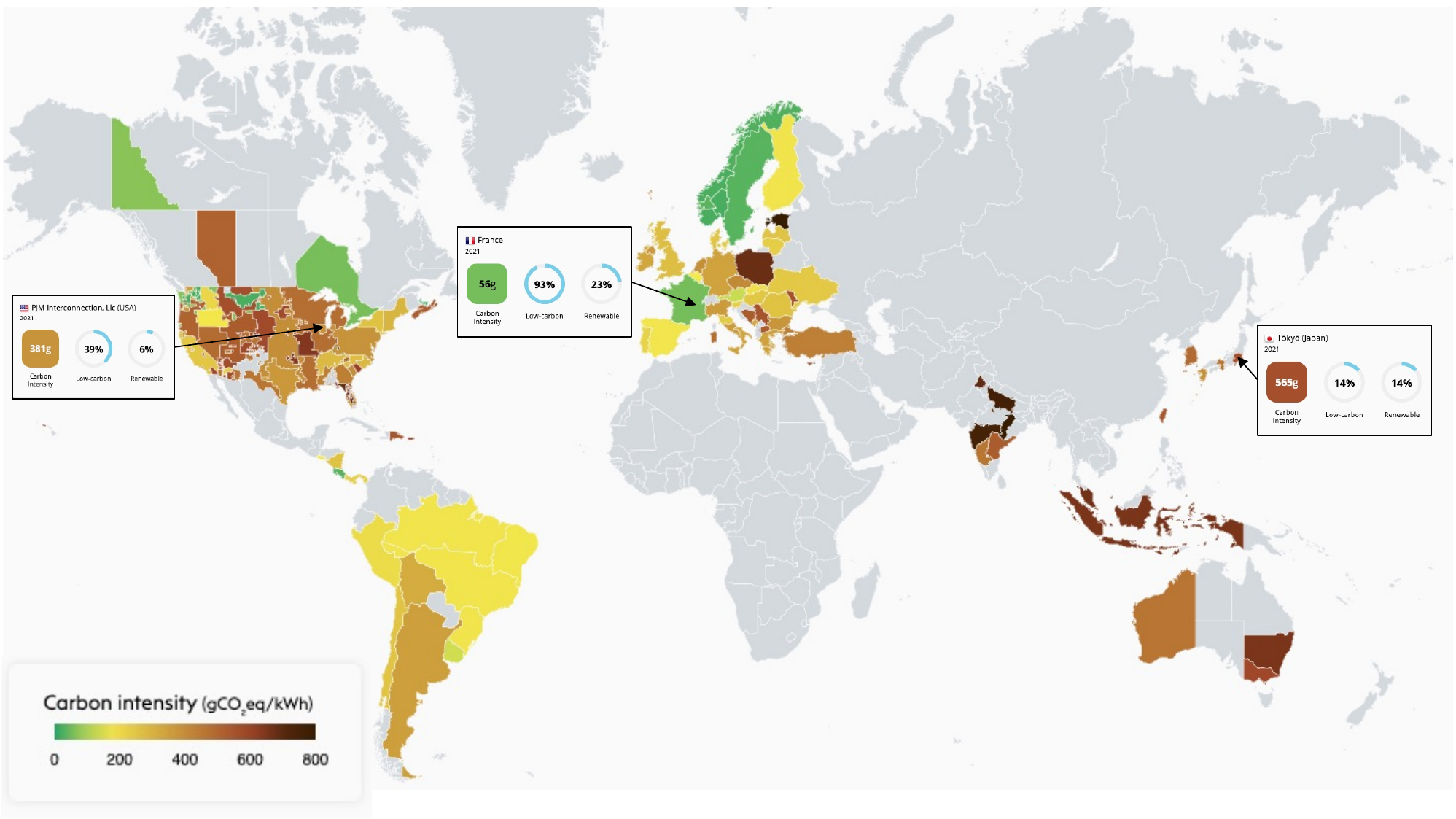}
\end{minipage}
\end{sideways}
\end{figure}

The carbon intensity for electricity production (in kg $\rm CO_2$ eq. / MWh) by the different countries can be obtained from Ref.~\cite{CO2}, at least for Japan, France and US Illinois, as shown in Fig.~\ref{fig:CO2} for 2021. Data from China were not made available to this application at the time of writing this note, and the specific location of CEPC is not known, but average data over the whole China in 2021 can be found at \url{https://lowcarbonpower.org/map-gCO2eq-kWh-iea}, still allowing a rough footprint estimate to be made. These data are displayed in the second row of Table~\ref{tab:CO2} for each of the five $\rm e^+e^-$ Higgs factory projects, under the assumptions that CLIC or FCC-ee would be operated at CERN, ILC at KEK, $\rm C^3$ at FNAL, and CEPC from somewhere in China. The carbon footprint of each Higgs boson produced is the product of the energy consumption per Higgs boson from Table~\ref{tab:Higgs} by the carbon intensity of the energy used by each Higgs factory, and is displayed in the last row of Table~\ref{tab:CO2}. 

\begin{table}[htbp]
\centering
\caption{For each of the Higgs factory projects (1st row): Assumed collider location (2nd row); Carbon intensity of electricity production in 2021 at each location, except for China where the carbon intensity is averaged over the whole 2021 (3rd row); Corresponding carbon footprint of each Higgs boson produced (last row).}
\label{tab:CO2}
\begin{center}
\begin{tabular}{@{}lccccc@{}}
\toprule
Higgs factory & CLIC & ILC & $\rm C^3$ & CEPC & FCC-ee\\
Operated from  & CERN & KEK & FNAL & China & CERN \\
\midrule
\begin{tabular}{l}
Carbon intensity\\ (kg $\rm CO_2$ eq.\,/\,MWh)
\end{tabular}
& 56 & 565 & 381 & 546 & 56 \\
\midrule
\begin{tabular}{l}
Carbon footprint per Higgs boson\\ 
(t $\rm CO_2$ eq.) 
\end{tabular}
& 0.8 & 9.4 & 6.8 & 2.2 & {\bf 0.17} \\
\botrule
\end{tabular}
\end{center}
\end{table}

Should it operate today, the Higgs factory with the least environmental impact, i.e., with the smallest carbon footprint for a given physics outcome, would therefore be FCC-ee operating at CERN, a factor 5 ahead of the next contender (CLIC). With 4 IPs, the FCC-ee carbon footprint would further reduce to 0.1\,t $\rm CO_2$ eq. for each Higgs boson produced, {\bf almost two orders of magnitude better than ILC operating at KEK}. 

These estimates will need to be revised regularly, as the energy carbonation levels might change and hopefully improve in developed/developing countries by the end of HL-LHC. For example, over the 2005-2019 period, the carbon intensity of the French economy fell by 32\% while GDP increased by 17\%. France plans to further reduce the carbon intensity of its electricity production by another 40\% by the end of the decade (and by a factor of almost three by 2050)~\cite{France}, by further developing its nuclear fleet and by increasing the production of renewable energies. A 450 B\$ Climate Bill has just been adopted by US Senate (at the beginning of August 2022) with a similar target for the decade to come. As of 23 July 2022, only 6 out of 17 major nuclear power plants operate in Japan~\cite{Japan}, while the 11 others are pending restart. Together with the development of renewable energies in Japan, this may allow the ILC figures to reduce as well in similar proportions. Finally, 
the carbon intensity of China's electricity production was already reduced by 15\% between 2011 and 2021~\cite{China}, and this reduction is expected to amplify after 2030, to comply with the 2015 Paris agreements. 

\section{Conclusion}\label{sec13}

The discussions of the relative scientific merits of linear and circular Higgs factories (CLIC, ILC, $\rm C^3$, CEPC, and FCC-ee) have been extensive throughout the European Strategy Update and the Snowmass studies, with the various communities having agreed on standard candles of performance comparison. 
Given the increasing relevance of the environmental impact of future colliders, common meaningful criteria for the comparison between Higgs factories, which differ from those that have been tabulated so far (e.g., instantaneous power, annual energy consumption), have been proposed in this note, relating the scientific output to the energy consumption, or to the carbon footprint.  These criteria exhibit very significant differences between the five contenders and might therefore be a convincing tie-breaker in these times of global warming. 
\begin{enumerate}
\item Because of the much larger instantaneous luminosity and the possibility of operating four detectors, circular Higgs factories are found to be more energy efficient to produce Higgs bosons -- the raison-d'\^etre of a Higgs factory -- than their linear counterparts. Specifically, the energy consumed by FCC-ee to produce one Higgs boson is expected to be an order of magnitude smaller than by any of the proposed linear Higgs factories. This conclusion is valid independently of assumptions on the origin of electricity.
\vskip .2cm
\item Because the electricity used at CERN is mostly carbon-free, the carbon footprint of the FCC-ee operating as a Higgs factory with 4 IPs at CERN would be, if operated today, typically two orders of magnitude smaller than that of ILC/$\rm C^3$, a factor 20 smaller than that of CEPC, and almost one order of magnitude smaller than that of CLIC (which would use the same electricity production mix as FCC), for the same measure of scientific outcome. 
\end{enumerate}

\begin{figure}[htbp]%
\centering
\caption{Energy consumption (top) and carbon footprint (bottom) for the five Higgs factory projects (CLIC at 380\,GeV, ILC/$\rm C^3$ at 250\,GeV, and CEPC/FCC-ee at 240\,GeV), per Higgs boson produced, i.e., for an equivalent physics outcome. In these plots, FCC-ee is assumed to operate only two detectors. With four detectors, the FCC-ee estimators would be divided by a factor 1.7.} \label{fig:Conclusion}
\includegraphics[width=0.87\textwidth]{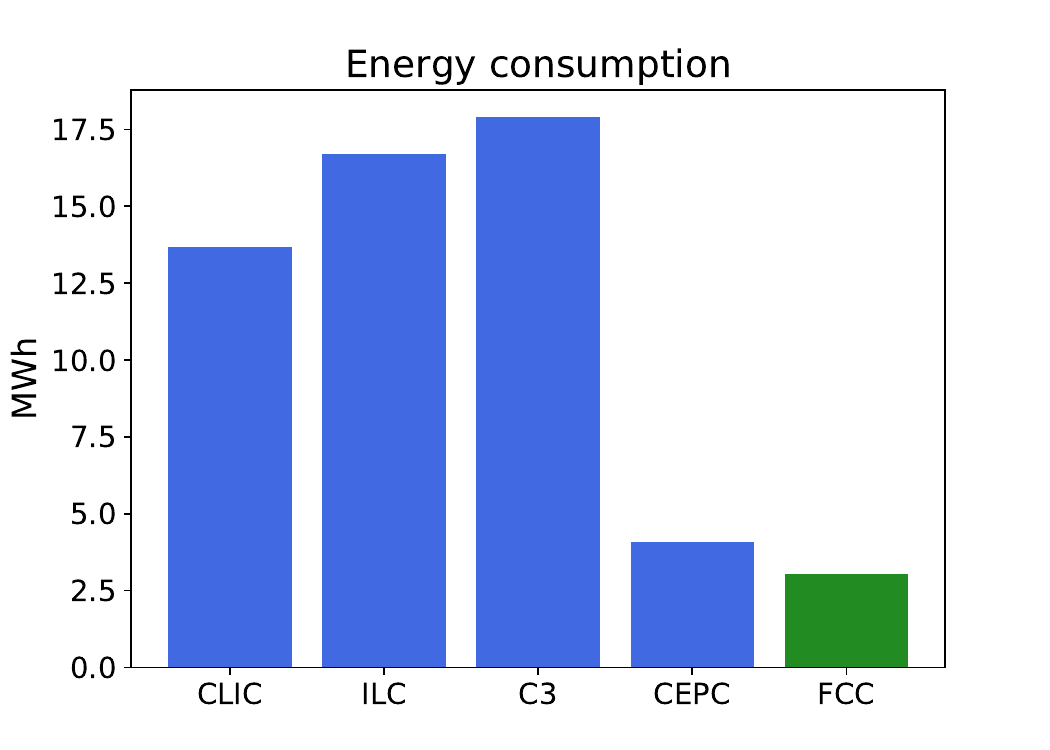}
\includegraphics[width=0.87\textwidth]{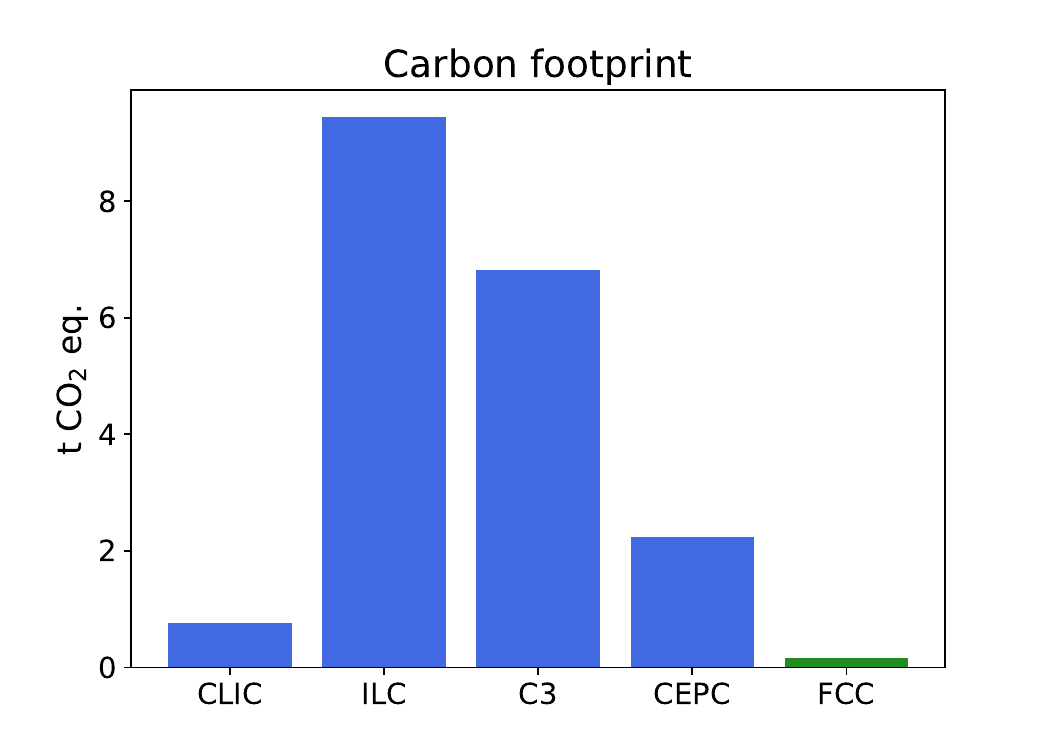}
\end{figure}

\noindent In short, and as illustrated in Fig.~\ref{fig:Conclusion}, {\bf FCC-ee is -- by very large factors -- the least disruptive in terms of environmental impact during operation}, among the ${\rm e^+e^-}$ candidate Higgs factories aimed at operating by the end of HL-LHC. 

Certainly, the numbers presented in this note are not devoid of uncertainty: the efficiency of real machines, i.e., the fraction of the available facility power that can be used; and the fraction of the time this power can be used; can only be known after the fact.As a result of the constant and worldwide ongoing efforts, the carbon intensity of electricity production is also expected to be much smaller by the time of the aspirational start of operations of the considered colliders, in the mid-2030 for ILC, $\rm C^3$ or CEPC, and even more so by mid/end-2040 for FCC-ee or CLIC. All the above ``predictions'' must therefore be read relative to each other and with some caution, at the 10-20\% level and perhaps more. Nevertheless the relative differences are well beyond these uncertainties. Meanwhile, common sustainability studies and R\&D activities are being pursued both in the laboratory and in industry to further improve the energy efficiency of these colliders, along the four lines of sobriety, efficiency, energy management and waste heat recovery. 

\backmatter

\bmhead{\large Discussion and supplementary information}

\begin{itemize}
    \item This note is part and parcel of a much more general enterprise, the ``FCC Feasibility Study''~\cite{FCC-FS-org:202106,FCC-FS-plans:202106}, approved and funded by the CERN Council for a period of five years (2021-2025), during which a high-level climate impact study is being carried out for the different phases of the project: tunnel excavation and infrastructure, installation and operation of FCC-ee, installation and operation of FCC-hh, decommissioning. The overall estimate of the FCC environmental impact will come from the compilation of expert studies from a large number of stakeholders with very different qualifications. 
    \vskip .2cm
    \item We (A.B. and P.J.) are particle physicists. We therefore focused on the part of the study we master most and where we can actually act towards reducing the environmental impact, namely the physics outcome of the proposed $\rm e^+e^-$ Higgs factories, e.g.  by raising awareness about the fact that some colliders could pollute up to 100 times less than others for a given physics outcome. 
    \vskip .1cm
    Specifically, we are proposing that future high-energy physics projects include not only the collider cost and power, but also its carbon footprint per physics outcome, and to use these data in the design and the choice of the next collider. Even more specifically, we propose -- and this is the least we can do as a community -- to maximise the physics outcome for a given carbon footprint (or, equivalently, minimise the carbon footprint for a given physics outcome). This requires a careful optimization of the collider design and of its operation model, and {\it in fine} the proper choice of the collider project. This proposed (r)evolution of the evaluation process places the environmental future of our planet as one of the top-level decision figures of merit. 
    \vskip .2cm
    \item All the facilities mentioned in this note require the construction of a new collider infrastructure, which adds to the carbon footprint from running the collider. (Two facilities, ILC and CEPC, would start from a green field, and thus require in addition the construction of a brand-new international particle physics centre.) Because the FCC-ee footprint in operation is so much smaller than that of the other contenders, it is important to make sure that the infrastructure, and particularly the tunnel construction, does not saturate the total carbon footprint. 
    \vskip .1cm
    As already mentioned, a high-level climate impact study of the civil engineering is being carried out in the context of the FCC feasibility study.
    Several estimates already exist in the literature for the 5.5-m-wide, 91-km-long FCC tunnel, should conventional construction methods be used. An experimental approach from existing road tunnel construction is followed in Ref.~\cite{RODRIGUEZ2021103704}, extrapolating to 250-300\,kt $\rm CO_2$ for the FCC tunnel, assuming that the footprint scales with the tunnel length and the square of its diameter. In a more recent contribution to Snowmass'21~\cite{Bloom:2022gux}, a footprint of 221\,kt $\rm CO_2$ is predicted from a bottom-up calculation driven by the tunnel parameters. Both estimates are consistent, and are comparable to the 170\,kt $\rm CO_2$ eq. from the three years of FCC-ee running at $\sqrt{s} = 240$\,GeV as a Higgs factory.
    Importantly, the tunnel construction footprint is also significantly smaller than the $\simeq 1.25$\,Mt $\rm CO_2$ eq. expected from the 15 years of the FCC-ee programme~\cite{Burnet}. The same tunnel may then serve a 100\,TeV hadron collider (and/or potentially a multi-TeV muon collider) for several additional decades. However narrow it may appear to an outsider to the field, the focus of this note may therefore well turn out to be the central aspect of the feasibility study in terms of environmental impact, i.e., that which would benefits most from further optimisation.
    \vskip .2cm
    \item In this note, the annual energy consumption of each Higgs factory includes only the collider consumption while in operation, including downtime. (Most of the operation parameters for $\rm C^3$ have been assumed to be identical to the corresponding ILC published numbers.) An alternative footprint estimate could be obtained by including the energy spent in beam commissioning, machine-development periods, technical stops, and shutdown periods. Such an exercise has been done, e.g., for FCC-ee in Ref.~\cite{Burnet} and for CLIC in Ref.~\cite{Brunner:2022usy}, which increases the annual energy consumption from 1.0 to 1.52\,TWh, and from 0.4 to 0.6\,TWh, respectively. Because similar factors would apply to the other Higgs factories as well, the conclusions of this note are largely unchanged with this more complete estimate. 
    \vskip .2cm
    \item It is often argued that the physics outcome of a Higgs factory does not depend solely on the number of Higgs bosons produced, but that longitudinal beam polarization (foreseen at linear colliders) is also to be taken into account in the energy-efficiency determination. Table~\ref{tab:pol} displays the projected statistical precision of the Higgs coupling measurements made with one million Higgs bosons, produced at $\sqrt{s} = 250$\,GeV either with or without longitudinal polarization, as extrapolated from Table XIX of Ref.~\cite{Bambade:2019fyw}. 
    \vskip .1cm
    It can be inferred from Table~\ref{tab:pol} that, while longitudinal beam polarization may increase the $\rm e^+e^- \to ZH$ production cross section -- by a factor already included in all estimates made so far in this note -- the projected uncertainties (i.e., the physics outcome) with a given number of Higgs bosons are mostly independent of whether beams are longitudinally polarized or not. The small excursions (within $\pm 10\%$) observed with an Effective Field Theory (EFT)~\cite{Grzadkowski:2010es} method entirely disappear~\cite{Blondel:2019yqr} when the couplings are determined in the $\kappa$ framework~\cite{LHCHiggsCrossSectionWorkingGroup:2012nn,Heinemeyer:2013tqa}. 

\begin{table}[htbp]
\centering
\caption{Projected statistical uncertainties (in \%) on the Higgs boson couplings with one million Higgs bosons produced in the $\rm e^+e^- \to ZH$ process at $\sqrt{s} = 250$\,GeV, obtained with the EFT method without (2nd row) and with (3rd row) longitudinal beam polarisation, as straightforwardly extrapolated from Table XIX of Ref.~\cite{Bambade:2019fyw}. The 2nd row includes an improvement of the uncertainties on precision electroweak observables by at least an order of magnitude with respect to the current world average, achievable at circular colliders with the detector calibration runs at $\sqrt{s} = m_{\rm Z}$ foreseen during the Higgs factory campaign. The 3rd row includes an improvement of the uncertainty on the electron electroweak couplings, achievable at linear colliders with the measurement of the $\rm e^+e^- \to \gamma \rm Z$ cross section with polarised beams at $\sqrt{s} = 250$\,GeV.}
\label{tab:pol}
\begin{center}
\begin{tabular}{@{}lccccc@{}}
\toprule
Higgs Coupling & ZZ/WW & bb & cc & gg & $\tau\tau$\\
\midrule
\begin{tabular}{l}
Precision w/o long. polarization (\%)
\end{tabular}
& 0.41 & 0.72 & 1.2 & 1.1 & 0.81 \\
\begin{tabular}{l}
Precision w/ long. polarization (\%)
\end{tabular}
& 0.36 & 0.71 & 1.3 & 1.2 & 0.79 \\
\botrule
\end{tabular}
\end{center}
\end{table}

    \item The current status of sustainability studies and related R\&D activities to further improve the energy efficiency and environmental impact of linear colliders at CERN, FNAL, or KEK, can be found in Ref.~\cite{Benny}. Many of the solutions presented therein are considered for circular colliders as well. 
    \vskip .1cm
    As a matter of fact, CERN  and all other candidate hosts can do a lot in the direction of reducing the carbon footprint estimated today: further investigate
    accelerator fundamental physics laws to optimise the collider specific luminosity (i.e., the luminosity for a given current); develop higher energy efficiency technology (e.g., more efficient RF power sources); generalisation of dissipated energy recovery to heat the neighbourhood (already at play with the LHC); think of alternative ways of energy storage (to be able to use renewable energies like solar energy or wind energy 24 hours a day); maximise synergies with developments of carbon-free energy production (e.g., nuclear fusion?); adapt electricity consumption with agility (e.g., operate the collider at full power when carbon-free energy is available, turn it off  otherwise); work on new ideas to transcend the limits of silicon for real and simulated data storage and analysis (e.g., quantum computing, optical computing, DNA data storage?); find ways of efficient international collaboration that minimise air-plane travel; etc. It is important to realise that all these efforts are highly incentive of innovative developments, which will not only be beneficial to energy consumption (and therefore to the related carbon footprint) at CERN or elsewhere, but they will also serve the society at large.
\end{itemize}


\bmhead{Acknowledgments}

Frank Zimmermann must be praised and acknowledged, for he has been advocating the FCC high efficiency per Higgs boson (and the even higher efficiency per Z boson!) for a long time.
Special thanks are also due to 
Panagiotis Charitos, 
Marcin Chrz\k{a}szcz, 
Mike Koratzinos, 
Joachim Kopp,
Michelangelo Mangano, 
Bill Murray,
Emmanuel Perez,
Vladimir Shiltsev
and Steinar Stapnes,
for their useful comments and suggestions during the preparation of this note. This work is partially funded from the European Union's Horizon 2020 research and innovation programme under grant agreement No. 951754.

\bibliography{sn-bibliography}

\section*{\small Data availability}
{\small \it \noindent Data supporting the findings of this study are available from the corresponding author upon request.}

\end{document}